\documentclass[aip,jcp,numerical,reprint]{revtex4-1}
\usepackage[version=3]{mhchem} % Formula subscripts using \ce{}
\usepackage{color}
\usepackage{graphicx}
\usepackage{threeparttable}
\usepackage{longtable}
\usepackage{enumerate}
\usepackage{amsfonts,amssymb}
\usepackage{amsmath}
\usepackage{stmaryrd}
\usepackage{dcolumn}
\usepackage{bm}
\usepackage{bbm}
\usepackage{algpseudocode}
\usepackage{algorithm}
\usepackage[all,cmtip]{xy}
\usepackage{url}

\bibpunct{}{}{,}{s}{}{\textsuperscript{,}}  % upper cite

\newcommand{\tr}{\mathrm{tr}}

\newcommand{\rvN}{\mathrm{vN}}

\begin{document}
%%% ----------------------------------------------------------------------

\title{
Expressibility of comb tensor network states (CTNS) for the P-cluster
and the FeMo-cofactor of nitrogenase
%polynuclear transition metal complexes:
%representation of selected configuration interaction wavefunctions
}
\author{Zhendong Li}\email{zhendongli@bnu.edu.cn}
\affiliation{Key Laboratory of Theoretical and Computational Photochemistry, Ministry of Education, College of Chemistry, Beijing Normal University, Beijing 100875, China}
\date{\today}

\begin{abstract}
Polynuclear transition metal complexes such as the P-cluster and the FeMo-cofactor of nitrogenase with
eight transition metal centers represent a great challenge for current electronic
structure methods. In this work, we initiated the use of comb tensor network states (CTNS),
whose underlying topology has a one-dimensional backbone and
several one-dimensional branches, as a many-body wavefunction ansatz
to tackle these challenging systems. As an important first step, we
explored the expressive power of CTNS with different underlying topologies.
To this end, we presented an algorithm to express a configuration interaction (CI) wavefunction into
CTNS based on the Schmidt decomposition.
The algorithm was illustrated for representing approximate CI wavefunctions
obtained from selected CI calculations
for the P-cluster and the FeMo-cofactor into CTNS with three chemically
meaningful comb structures, which successively group orbitals belonging to
the same atom into branches. The conventional matrix product states (MPS) representation was obtained
as a special case. We also discussed the insights gained from such decompositions,
which shed some light on the future developments of efficient numerical tools for
polynuclear transition metal complexes.
\end{abstract}
\maketitle

\newpage
\section{Introduction}
Polynuclear transition metal complexes have many fascinating properties. They
can be found as single molecular magnets\cite{caneschi1991alternating} or catalysts for difficult reactions
such as dinitrogen fixation\cite{falcone2017nitrogen,seefeldt2020reduction}. A prominent example in the later category
is the nitrogenase (Fig. \ref{fig:nitrogenase}), which contains
two pairs of metalloclusters. The P-cluster has an \ce{[Fe8S7]} core\cite{peters1997redox},
and is postulated as the intermediate for electron transfer
during the nitrogen fixation\cite{seefeldt2012electron}, which happens on
the FeMo-cofactor (FeMoco or M-cluster) with an \ce{[MoFe7S8C]} core\cite{spatzal2011evidence}.
Understanding their electronic structures
is the very first step towards unveiling the unusual properties of these complexes.
Unfortunately, the computational cost for an accurate
full configuration interaction (FCI) description
increases exponentially as the number of transition metal centers increases.
This is the case even for a minimal active space based theoretical model,
where only the $3d$ ($4d$) orbitals of the Fe (Mo) atoms and
the $3p$ orbitals of the S atoms are considered.
Our previous work\cite{li2019electronicPcluster} showed that for the resting state of the P-cluster (P$^{\mathrm{N}}$),
a chemically meaningful active space comprising 114 electrons and 73 spatial orbitals
(denoted by CAS(114e,73o) later for brevity) can lead to a
Hilbert space with $2.8\times 10^{31}$ determinants for the spin projection $M=0$.
For the ground state of the FeMoco with $S=3/2$,
while a CAS(54e,54o) model was previously suggested\cite{reiher2017elucidating},
it is recently shown that to correctly capture the open-shell characters of the
transition metal centers, a CAS(113e,76o) model is necessary\cite{li2019electronicFeMoco},
which leads to a Hilbert space with dimension about $3.6\times 10^{35}$.
These two complexes represent the most challenging metalloclusters
in nature for the present electronic structure theories,
which even makes them a potential target for quantum computers as
killer applications\cite{reiher2017elucidating,berry2019qubitization}.

The density matrix renormalization group (DMRG) algorithm\cite{white_density_1992,white_density-matrix_1993,schollwock_density-matrix_2005,schollwock_density-matrix_2011}, first developed
in condensed matter physics for strongly correlated models, has emerged
as a powerful tool for strongly correlated molecular systems\cite{chan_density_2011,Reiher2011,Sebastian2014,Legeza2015,Yanai2015,olivares-amaya_ab-initio_2015,baiardi2020density}.
With matrix product states (MPS)\cite{MPS1995} as
the underlying variational wavefunction ansatz in DMRG, the size of the
variational space is controlled by a single parameter $D$,
commonly referred as the bond dimension.
For one-dimensional systems where the entanglement is usually limited,
the ground state can be well captured by a finite $D$ independent of the
system size\cite{eisert2010}. For higher-dimensional systems, such as the
case for polynuclear transition metal clusters,
since the computational scaling of DMRG is $O(K^3D^3+K^4D^2)$
with $K$ being the number of spatial orbitals, which
is relatively low with respect to $D$, an accurate
description may still be obtained by increasing $D$ to
$O(10^3\text{-}10^4)$.
This has been shown for systems as complex
as the oxygen-evolving complex\cite{kurashige2013entangled} \ce{[Mn4CaO5]} and
the iron-sulfur clusters with \ce{[Fe2S2]} and \ce{[Fe4S4]} cores\cite{sharma_low-energy_2014},
by developing efficient
\emph{ab initio} DMRG algorithm using
symmetries and parallelizations\cite{white_ab_1999,chan_highly_2002,chan_algorithm_2004,sharma_spin-adapted_2012}.
For the P-cluster of nitrogenase, the first \emph{ab initio} investigation
has recently been accomplished\cite{li2019electronicPcluster} by integrating several state-of-the-art techniques,
including spin-projected DMRG\cite{li2017spin} for generating chemically meaningful initial MPS
and spin-adapted DMRG\cite{sharma_spin-adapted_2012} for efficiently approaching convergence
with large $D$. Such large-scale applications, however,
require a large amount of computational resources.
To make the \emph{ab initio} calculation of
polynuclear transition metal complexes as large as
the P-cluster and the FeMoco become
routine applications, new innovations
in theories and algorithms are necessary.

In this work, we initiate the use of comb tensor network states\cite{chepiga2019comb} (CTNS),
whose underlying topology has a one-dimensional backbone and
several one-dimensional branches, as a wavefunction ansatz
for these challenging systems. As the important first step,
we investigate the expressive power of CTNS with different underlying topologies.
To this end, based on the Schmidt decomposition, we present an algorithm to
represent an arbitrary configuration interaction (CI) wavefunction by CTNS.
Note that the MPS representation can be obtained as a special case.
As a byproduct, it also allows us to compute the entanglement
entropy for CI wavefunctions, which can provide some insights
into the electronic structures of these complexes.

The remaining part of the paper is organized as follows.
In Sec. \ref{sec:theory}, we present the theory of
CTNS and the algorithm for exactly representing
a CI wavefunction by CTNS.
In Sec. \ref{sec:results}, we numerically illustrate the algorithm for transforming
CI wavefunctions obtained from selected CI (SCI) calculations for the P-cluster and the FeMo-cofactor
into CTNS with several chemically meaningful comb structures. Finally, the conclusion
and outlook are drawn in Sec. \ref{sec:conclusion}.

\section{Theory and algorithm}\label{sec:theory}
\subsection{Comb tensor network states (CTNS)}
Tensor network states can be viewed as special parameterizations of the
FCI wavefunction $|\Psi_{\mathrm{FCI}}\rangle$ in the occupation number representation,
\begin{eqnarray}
|\Psi_{\mathrm{FCI}}\rangle &=& \sum_{\{n_k\}}\Psi^{n_1\cdots n_K}|n_1\cdots n_K\rangle,\label{eq:FCI}\\
\Psi^{n_1\cdots n_K} &=& \tr(\prod_k \mathbf{T}^{n_k}[k]).\label{eq:TNS}
\end{eqnarray}
where $K$ denotes the number of spatial orbitals and  $|n_k\rangle\triangleq |n_{k\alpha}n_{k\beta}\rangle\in\{|00\rangle,|01\rangle,|10\rangle,|11\rangle\}$.
The symbol $\mathbf{T}^{n_k}[k]$ represents a tensor $T^{n_k}_{\alpha_k\beta_k\gamma_k\cdots \lambda_k}[k]$,
where $n_k$ is often referred as physical indices, and
indices like $\alpha_k$ referred as virtual indices are all contracted in Eq. \eqref{eq:TNS} as
indicated by the trace notation $\tr(\cdots)$.
The order (or rank) of $\mathbf{T}^{n_k}[k]$ depends on the specific wavefunction ansatz.
For MPS, Eq. \eqref{eq:TNS} has a chain structure,
\begin{eqnarray}
\Psi^{n_1\cdots n_K} = \sum_{\{\alpha_k\}} T^{n_1}_{\alpha_1}[1]
T^{n_2}_{\alpha_1\alpha_2}[2]\cdots T^{n_K}_{\alpha_{K-1}}[K],\label{eq:MPS}
\end{eqnarray}
where the boundary tensors are vectors $T^{n_1}_{\alpha_1}[1]$ (or $T^{n_K}_{\alpha_{K-1}}[K]$) for given $n_1$
(or $n_K$) and other tensors in the middle are matrices $T^{n_k}_{\alpha_{k-1}\alpha_k}[k]$ for given $n_k$.
Assuming the dimensions of $\alpha_k$ take the same value $D$, the number of variational parameters
in MPS \eqref{eq:MPS} is $O(KD^2)$. To faithfully represent a generic FCI state \eqref{eq:FCI},
$D$ goes as $O(4^{K/2})$, that is, exponential in $K$. However, the power of TNS
approach is that by properly choosing wavefunction ansatz in Eq. \eqref{eq:TNS},
a good approximation can usually be obtained for the low-energy states with $D$ increasing
mildly with $K$ for quantum chemistry problems.

In this work, we investigate a special class of TNS - comb TNS (CTNS),
whose underlying topology has a comb structure,
which is composed of a one-dimensional backbone and several one-dimensional branches.
It has recently been used for studying lattice models in condense matter physics such as the
spin-1/2 Heisenberg model\cite{chepiga2019comb}. Here we use it for quantum chemistry problems, in particular,
the electronic structures of polynuclear transition metal compounds. The motivations for
investigating this ansatz are twofold. First, it embodies the chemical intuition that
strongly correlated orbitals within each atom need to be first grouped together. Second,
its computational complexity is close to MPS and
lower than the generic acyclic (loop-free) TNS - tree TNS (TTNS)\cite{shi2006classical,murg2010simulating,
nakatani2013efficient,murg2015tree}.
The compromise between computational
complexities and expressive powers may leave some room for finding
more efficient ansatz beyond MPS for quantum chemistry problems.

\begin{figure*}
  \begin{tabular}{ccc}
  \includegraphics[width=0.4\textwidth]{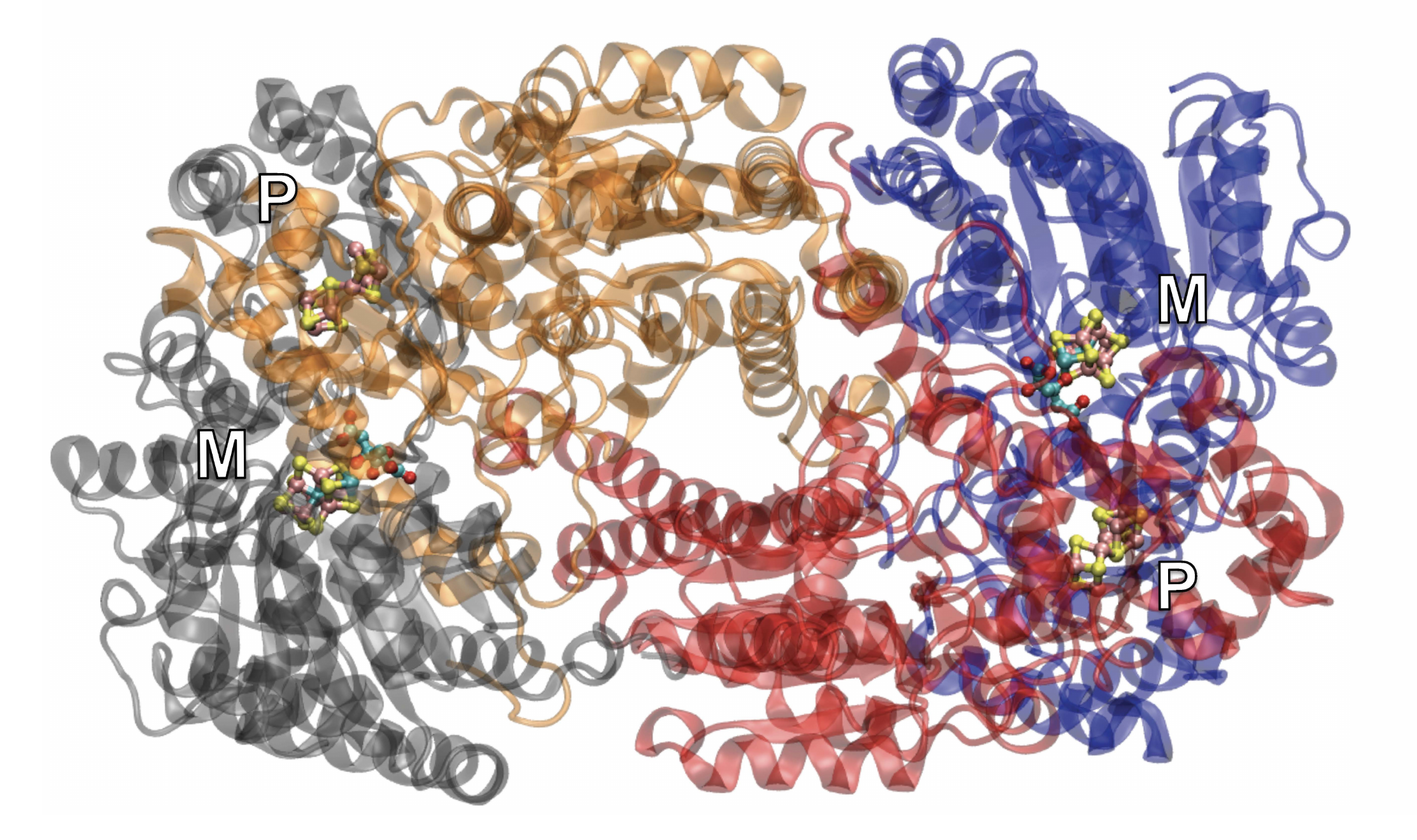} &
  \includegraphics[width=0.25\textwidth]{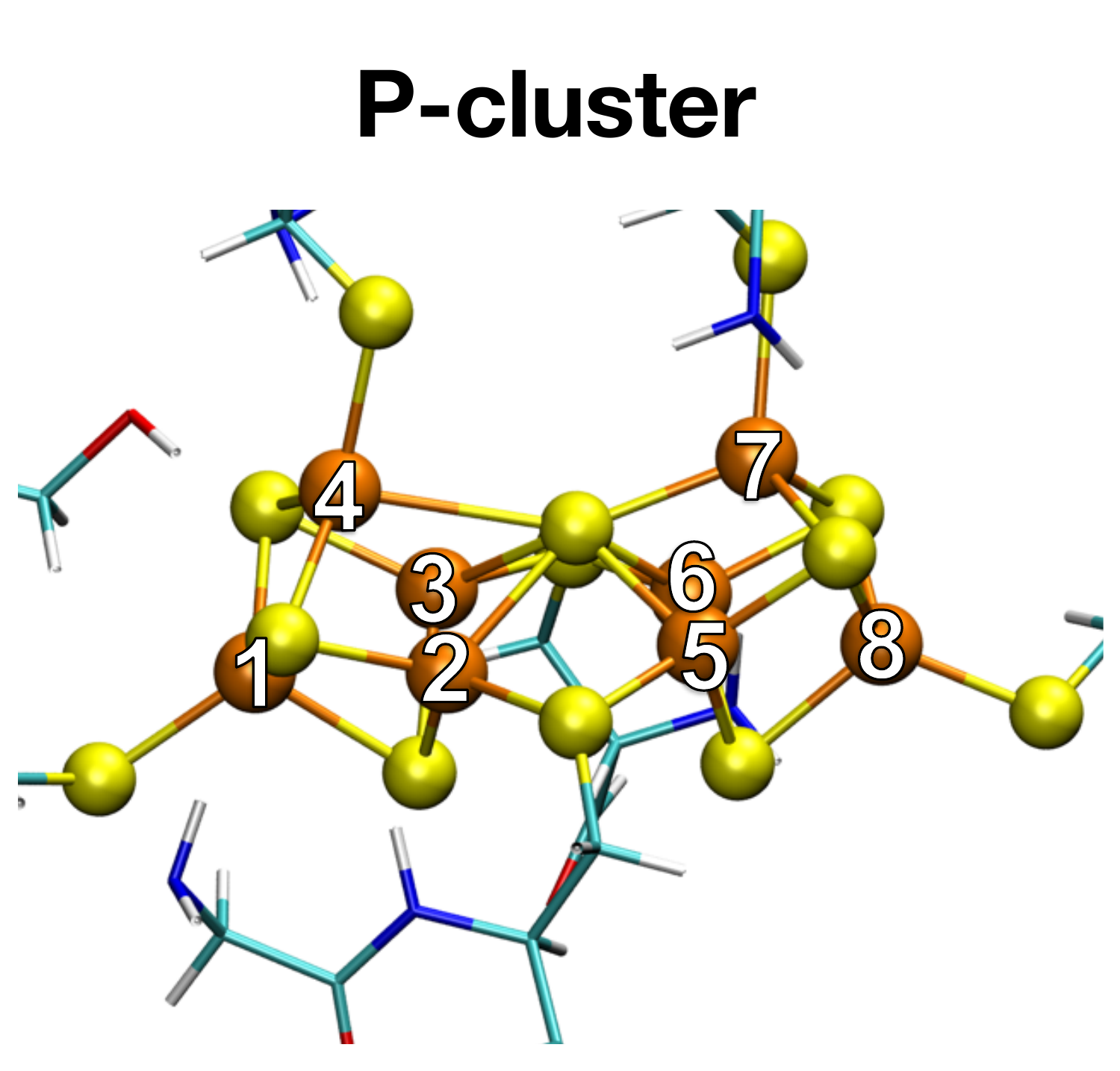} &
  \includegraphics[width=0.25\textwidth]{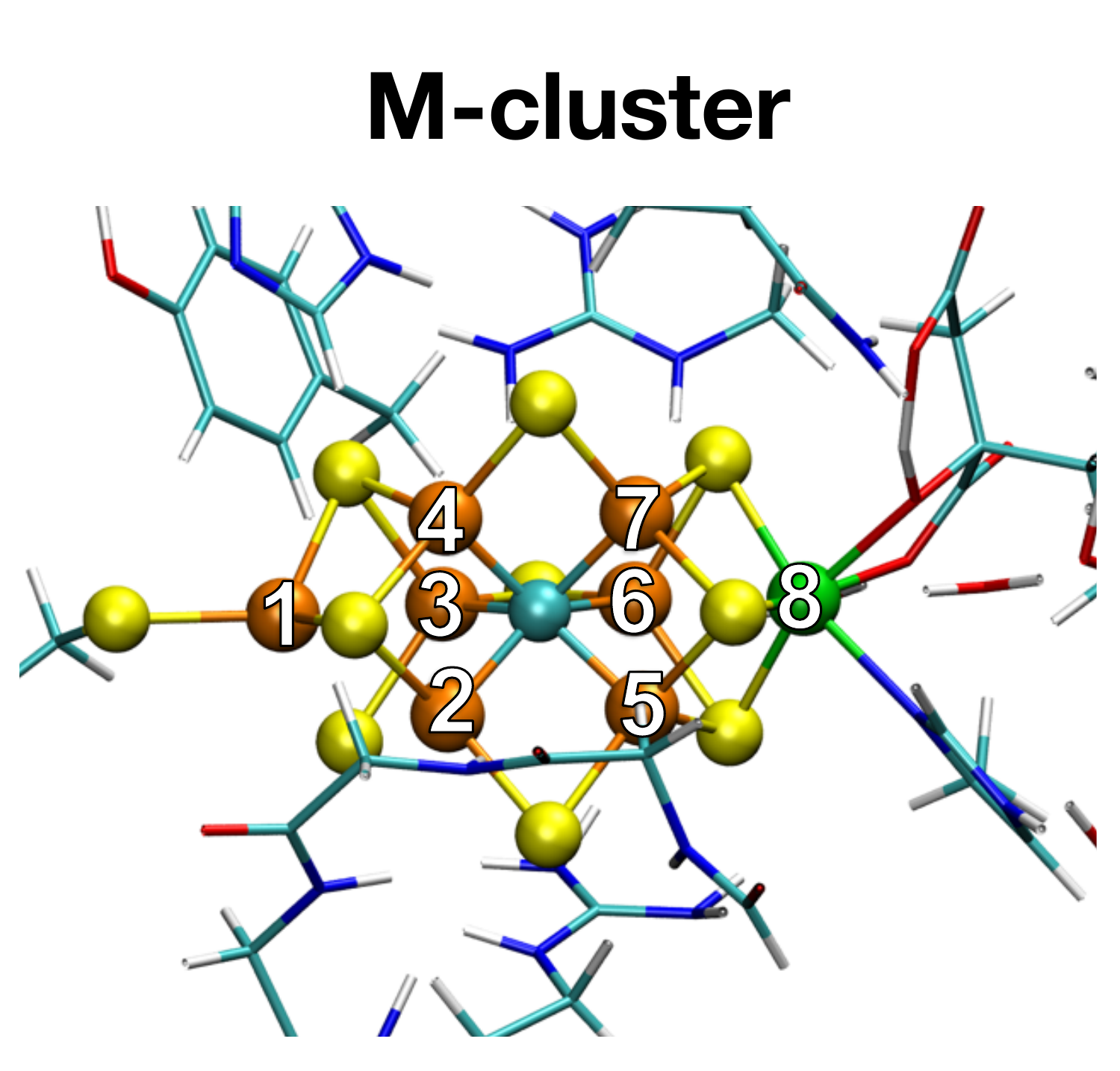} \\
  \multicolumn{3}{c}{(a) P-cluster and FeMoco (M-cluster) in nitrogenase} \\
  \multicolumn{3}{c}{\includegraphics[width=0.9\textwidth]{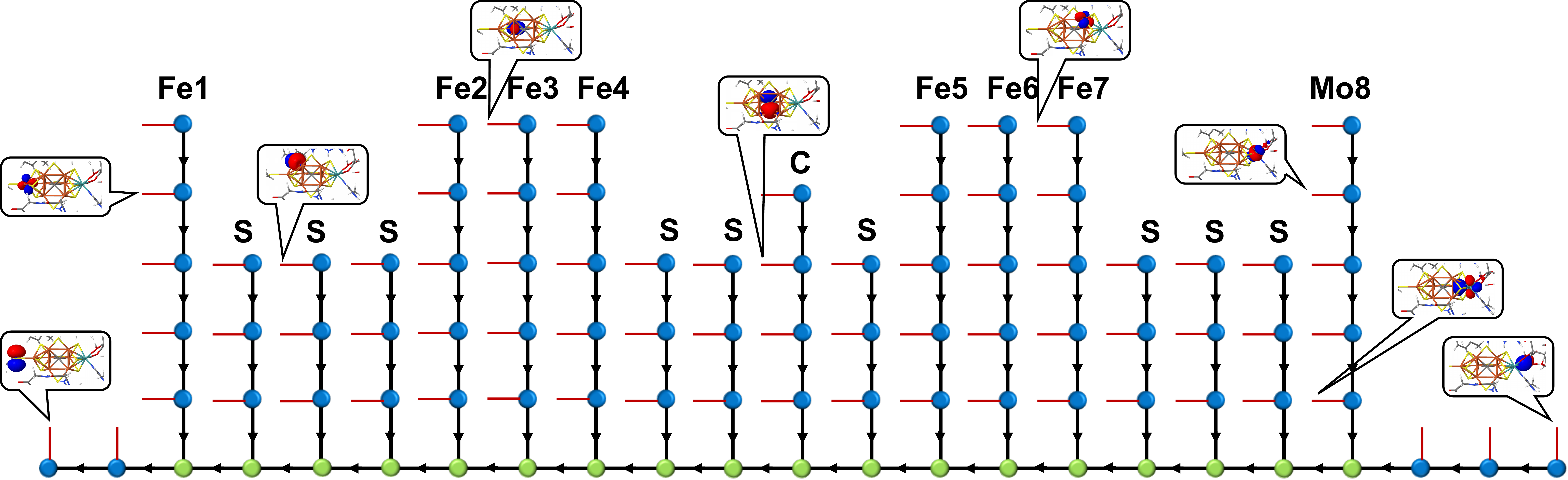}} \\
  \multicolumn{3}{c}{(b) Right canonical form of a CTNS for the FeMoco}
  \end{tabular}
  \caption{(a) The P-cluster and the FeMo-cofactor (M-cluster) in nitrogenase (PDB ID: 3U7Q).
  Color legend: Fe, orange; Mo, green; S, yellow; C, cyan; O, red; N, blue; H, white.
  The labels in the two complexes index the Fe/Mo atoms in the later figures.
  (b) The right canonical form of an CTNS for the active space model [CAS(113e,76o)]
  of the FeMoco. The sites in blue represent physical sites associated with spatial orbitals,
  while the sites in green represent internal sites without physical index (red lines).
  Some selected molecular orbitals are also illustrated.}\label{fig:nitrogenase}
\end{figure*}

An example of CTNS is
shown in Fig. \ref{fig:nitrogenase}(b) for the CAS(113e,76o) model of the FeMoco\cite{li2019electronicFeMoco}.
Each blue dot represents a physical site $\mathbf{T}^{n_k}[k]$ with the red line
representing the physical index $n_k$. The black lines between two tensors represent virtual indices
that are contracted. Slightly different from Eq. \eqref{eq:TNS}, we introduced
a set of internal tensors without physical index, see green dots in Fig. \ref{fig:nitrogenase}(b).
There are two motivations for introducing them.
Physically, they correspond to a coarse-graining operation which combines
the states of two branches ($V_1$ and $V_2$) into a reduced set of states in $V_1\otimes V_2$,
\begin{eqnarray}
|\alpha_{12}\rangle = \sum_{\alpha_1\alpha_2}
|\alpha_1\rangle|\alpha_2\rangle W_{\alpha_1\alpha_2,\alpha_{12}},\label{eq:CG}
\end{eqnarray}
where $|\alpha_1\rangle\in V_1$, $|\alpha_2\rangle\in V_2$, and $W_{\alpha_1\alpha_2,\alpha_{12}}$
is the coarse-graining transformation, which is also referred as isometry\cite{vidal2008class}
if $\mathbf{W}^\dagger\mathbf{W}=\mathbf{I}$.
Computationally, introducing these internal sites reduces the complexity
of TNS to those with only rank-3 tensors. It can be seen that if they are contracted with the connected
sites on the branches, the resulting tensors will become a rank-4 tensor,
similar to that in the generic TTNS. Such internal tensors are also
essential in the ansatz named three-legged TTNS (T3NS)\cite{gunst2018t3ns,gunst2019three}.
The difference between T3NS and CTNS is that CTNS is closer to MPS by design and
the two internal sites are allowed to be adjacent to each other, while
T3NS is derived from TTNS by inserting internal sites in a way that
internal sites are interleaved by physical sites in order to lower the computational complexity.
Furthermore, in connection to the multilayer multiconfiguration time-dependent Hartree theory (ML-MCTDH)\cite{MLMCTDH2009},
which employs a hierarchical tree TNS as ansatz\cite{hackbusch2014tensor,larsson2019computing},
we note that while MPS can be viewed as an unbalanced hierarchical binary tree,
the generic CTNS is more balanced.

Unlike MPS, whose underlying chain topology is unique, there can be different
topologies for CTNS. In the applications to lattice systems, the topology
is commonly determined by the underlying lattice structure, and the branches
can be quite long\cite{chepiga2019comb}. In contrast, motivated by chemical intuitions for
polynuclear transition metal compounds, we focus on comb topologies with relatively short branches.
Specifically, we will investigate three kinds of chemically meaningful topologies:

(1) Topology A is just MPS, which is a special CTNS without branches
or equivalently with branches of length one if internal sites are used.
This will serve as the reference for comparison.

(2) Topology B groups the $d$ orbitals within each transition metal atom (see Figs.
\ref{fig:pn} and \ref{fig:femoco}), leaving those
active orbitals of sulfur or carbon atom in the MPS-like backbone.

(3) Topology C further groups the active orbitals of sulfur
or carbon atom, as shown in Fig.
\ref{fig:nitrogenase}.

The expressive power of CTNS with
these three topologies will be compared
for representing CI wavefunctions
of the P-cluster and the FeMoco in the following sections.
Before we discuss the algorithm for representing CI wavefunctions
by CTNS. We mentioned that transforming an MPS
to CTNS is possible using an algorithm\cite{chepiga2019comb}
with successive contractions and singular value decompositions (SVD)
to shift virtual bonds.

\subsection{CTNS representation of CI wavefunctions}
We introduce an algorithm to represent an arbitrary CI wavefunction by CTNS
in two steps. In fact, this algorithm works for any loop-free TNS.
The first step is to compute all renormalized basis from
the CI wavefunction via the Schmidt decomposition.
The critical feature of acyclic TNS is that removing
a virtual bond leads to a bipartition of the
TNS into two parts, each with its own underlying
physical degrees of freedoms. We denote
the occupation basis of one space by $\{|n_l\rangle\}\triangleq
\{|n_{l_1}\cdots n_{l_m}\rangle\}$,
and that of the remaining space by
$\{|n_r\rangle\}\triangleq
\{|n_{r_1}\cdots n_{r_{K-m}}\rangle\}$.
Then, the CI wavefunction \eqref{eq:FCI} can be
rewritten as
\begin{eqnarray}
|\Psi_{\mathrm{CI}}\rangle = \sum_{lr}|n_ln_r\rangle \Psi^{lr}.\label{eq:CIlr}
\end{eqnarray}
Using the SVD of the matrix
\begin{eqnarray}
\Psi^{lr}=
(U\sigma V^\dagger)_{lr}=
\sum_{\alpha} U_{l\alpha} \sigma_\alpha V^{*}_{r\alpha},
\end{eqnarray}
the Schmidt decomposition of the CI wavefunction can be obtained as
\begin{eqnarray}
|\Psi_{\mathrm{CI}}\rangle = \sum_\alpha
|u_\alpha v_\alpha\rangle \sigma_\alpha,\label{eq:Schmidt}
\end{eqnarray}
where $\{|u_\alpha\rangle\}$ and $\{|v_\alpha\rangle\}$ form compressed
orthonormal states in the two spaces, respectively,
\begin{eqnarray}
|u_\alpha\rangle &=& \sum_l |n_l\rangle U_{l\alpha},\\
|v_\alpha\rangle &=& \sum_r |n_r\rangle V_{r\alpha}^*.\label{eq:RenormStates}
\end{eqnarray}
From Eq. \eqref{eq:Schmidt}, the bipartite entanglement
entropy, referred as von Neumann entropy $S_{\rvN}$, can be computed as
\begin{eqnarray}
S_{\rvN}=-\sum_i\lambda_i \log_2\lambda_i,\quad
\lambda_i = \sigma_i^2.
\end{eqnarray}
The nonnegative number $S_{\rvN}$ measures to what extent
the two subsystems are entangled with each other in the state $|\Psi_{\mathrm{CI}}\rangle$.
Clearly, if $|\Psi_{\mathrm{CI}}\rangle=|u\rangle|v\rangle$ is a product state,
then $S_{\rvN}$ achieves its minimal value zero.
This decomposition can be performed for each virtual bond of TNS,
and the obtained renormalized states are stored.

The second step is to construct the TNS representation with the
obtained renormalized states. This can be done by
choosing either the set $\{|u_\alpha\rangle\}$ or
$\{|v_\alpha\rangle\}$ on each bond. Graphically,
the choice corresponds to assign a direction to
the specific bond in TNS. In Fig. \ref{fig:nitrogenase}(b),
we illustrate a particular choice adopted in this work, referred
as the \emph{right canonical form}, where at each bond the set of renormalized states $\{|v_\alpha\rangle\}$ \eqref{eq:RenormStates}
is chosen. Then, physical and internal sites can be constructed from
\begin{eqnarray}
T^{n_k}_{\alpha_l\alpha_r}[k]=\langle n_k v_{\alpha_r}|v_{\alpha_l}\rangle,\label{eq:ovlpTk}\\
W_{\alpha_l\alpha_c\alpha_r}=\langle v_{\alpha_c} v_{\alpha_r}|v_{\alpha_l}\rangle,\label{eq:ovlpW}
\end{eqnarray}
respectively, where $v_{\alpha_c}$ (or $n_k$), $v_{\alpha_l}$, and $v_{\alpha_r}$ represent
the states on the respective central, left, and right bonds of
a tensor. The overlaps on the right hand sides of Eqs. \eqref{eq:ovlpTk} and \eqref{eq:ovlpW}
can be computed using the definition of renormalized states \eqref{eq:RenormStates}.
The ordering of indices for tensors on the left hand sides is
not important, as long as a consistent convention is used
in performing contractions.

Using the above two-step algorithm, we are able to represent
an arbitrary CI wavefunction by CTNS. Generalizing this algorithm to
represent multiple CI wavefunctions $\{|\Psi_i\rangle\}$ simultaneously
is straightforward, by using either SVD for an expanded coefficient matrix\cite{hubig2015strictly,larsson2019computing}
$\tilde{\Psi}^{il,r}\triangleq \Psi^{lr}_i$ or diagonalization of the
state-averaged reduced density matrix $\rho=\sum_i\Psi^T_i\Psi^*_i$
to define the renormalized basis $\{|v_\alpha\rangle\}$.
Furthermore, if $|\Psi_i\rangle$ is an eigenfunction of the total particle number operator
$\hat{N}$ and the spin projection operator $\hat{S}_z$, then
the reduced density matrix $\rho$ will be block-diagonal, with each
block corresponding to a definite particle number and spin projection.
Hence, the obtained renormalized states will also be eigenfunctions
of $\hat{N}$ and $\hat{S}_z$.

This algorithm can be applied for two purposes. It can be used to produce a good
initial CTNS for variational optimization,
which will be the subject of our future work.
In this work, we focus on using this algorithm as
a tool to analyze the expressive power of CTNS.
To this end, we will use a tight truncation threshold
in the Schmidt decomposition \eqref{eq:Schmidt}
such that the obtained CTNS is a faithful representation
of the original CI wavefunction. The overlap
between the CI wavefunction and the obtained CTNS
can be computed to ensure this. Then, we
compare the bond dimensions, which are the
dimensions of the resulting renormalized states,
and the entanglement entropies $S_{\rvN}$
for CTNS with different topologies.

\subsection{Implementation and computational details}
The above algorithm was implemented into an in-house program named \textsc{Focus} in C++.
Since FCI is not feasible for the P-cluster and the FeMoco, we used CI wavefunctions
obtained from selected CI (SCI) calculations as representatives
to investigate the expressive power of CTNS with different topologies.
To this end, we implemented the heat-bath CI algorithm\cite{holmes2016heat,sharma2017semistochastic,holmes2017excited},
which allows a fast exploration of the Hilbert space. One technical point deserves mentioning
is that since CI algorithms usually work with $\alpha$-string and $\beta$-string\cite{knowles1984new},
a transformation step needs to be carried out to convert the basis
into the occupation number representation \eqref{eq:TNS}.
Specifically, the necessary phase change can be derived as
\begin{eqnarray}
&&|n_{1\alpha},\cdots,n_{K\alpha},n_{1\beta},\cdots,n_{K\beta}\rangle\nonumber\\
&=&|n_{1\alpha},n_{1\beta},\cdots,n_{K\alpha},n_{K\beta}\rangle
(-1)^{
\sum_{i=1}^{K-1}\sum_{j=i+1}^{K}n_{j\alpha}n_{i\beta}
},
\end{eqnarray}
where $n_{k\alpha}$ and $n_{k\beta}$ are the occupation numbers of spin orbitals.

For each complex, we carried out six SCI iterations to generate a representative multi-determinant
CI wavefunction, where in each iteration the variational subspace grows following the
criteria of heat-bath CI\cite{holmes2016heat}. Specifically,
a new determinant $|A\rangle\notin\mathcal{V}$
not belonging to the current variational space $\mathcal{V}=\{|I\rangle\}$ of determinants
is selected if $\max_{I\in\mathcal{V}}|\langle A|H|I\rangle c_I| \ge \epsilon_1$ with
$\epsilon_1=10^{-3}$. This procedure generates
SCI wavefunctions with 19338 and 108393 determinants
for the P-cluster and the FeMoco using
the previously reported active space models\cite{li2019electronicPcluster,li2019electronicFeMoco}
with localized molecular orbitals, respectively.
The molecular integrals generated using \textsc{Pyscf}\cite{sun2018pyscf}
are available from the online repositories\cite{linkToFCIDUMPpclusters,linkToFCIDUMPfemoco}.
In the Appendix, we documented the details of the three topologies of CTNS for both clusters.

\section{Results and discussion}\label{sec:results}
\begin{figure*}
  \begin{tabular}{cc}
  \multicolumn{2}{c}{\includegraphics[width=0.8\textwidth]{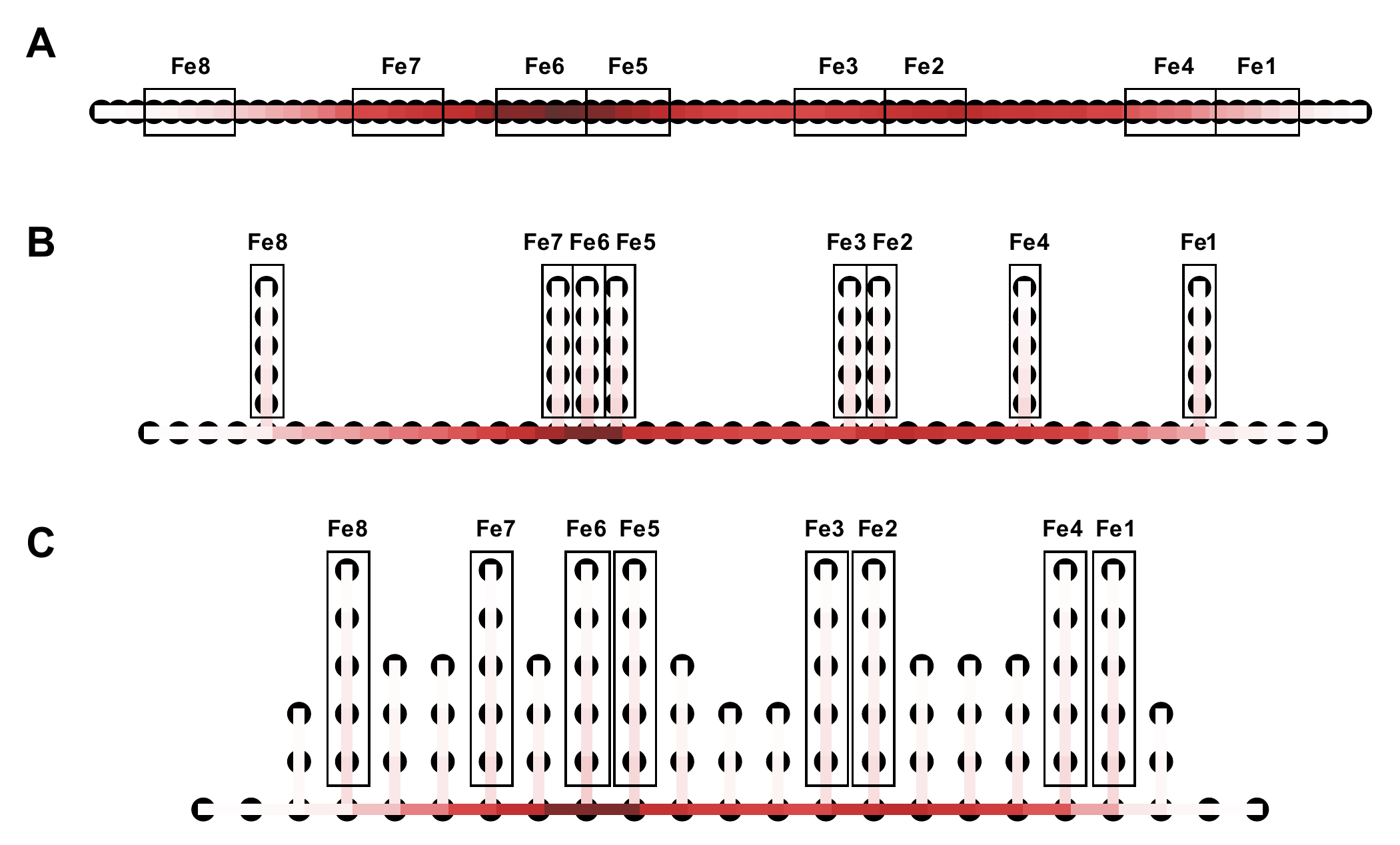}} \\
  \multicolumn{2}{c}{(a) three topologies of CTNS}\\
  \includegraphics[height=0.25\textheight]{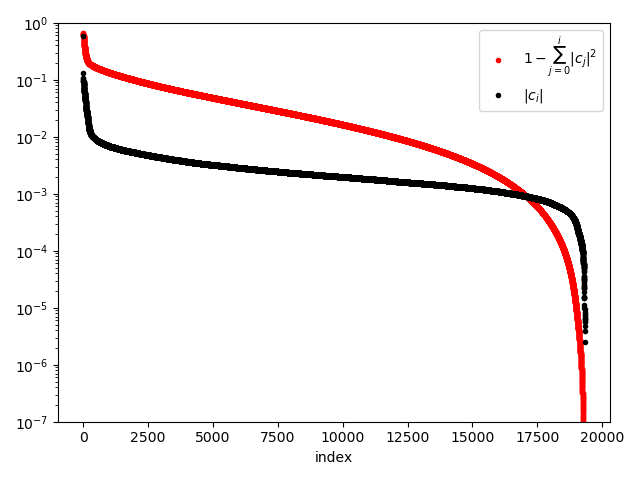} &
  \includegraphics[height=0.25\textheight]{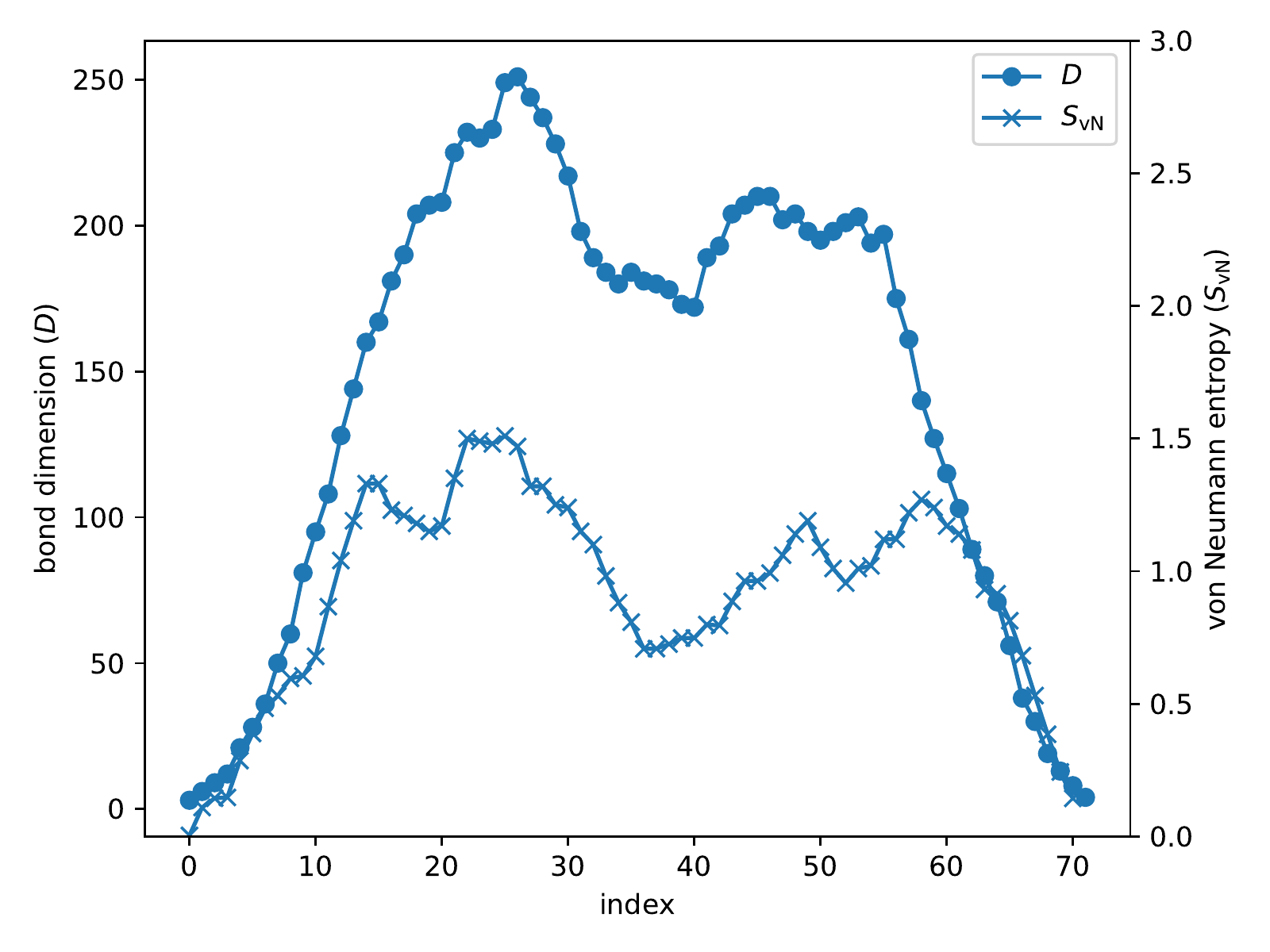} \\
  (b) magnitudes of SCI coefficients & (c) $D$ and $S_{\rvN}$ of MPS (topology A) \\
  \includegraphics[height=0.25\textheight]{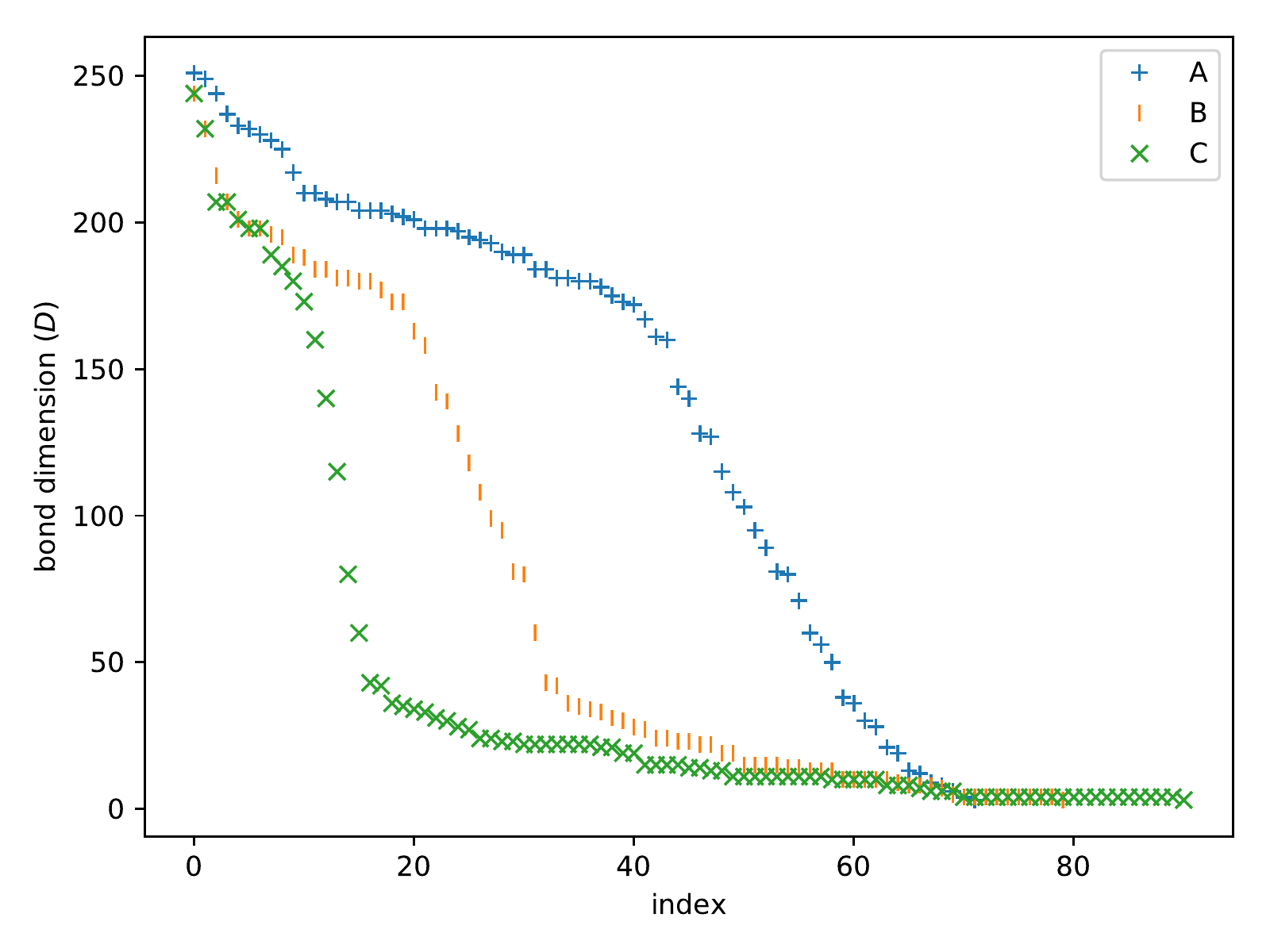} &
  \includegraphics[height=0.25\textheight]{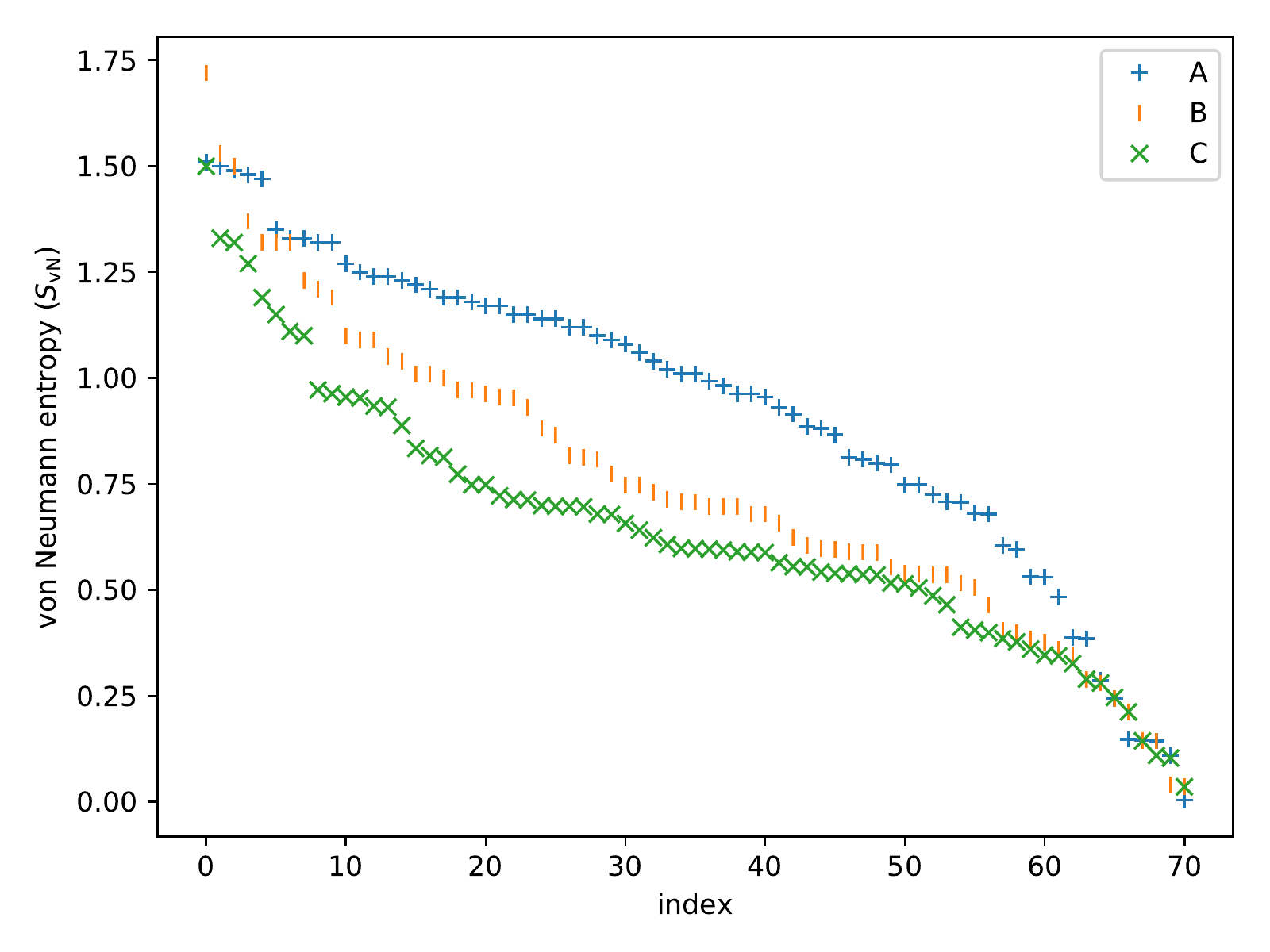} \\
  (d) sorted $D$ for three CTNS & (e) sorted $S_{\rvN}$
  \end{tabular}
  \caption{CTNS representations with different topologies of the SCI wavefunction ($M=0$) obtained with $\epsilon_1=10^{-3}$
  for the P-cluster in the active space CAS(114e,73o). (a) Three topologies of CTNS. Each black dot represents a tensor in CTNS, where the physical indices are omitted for simplicity. Darker color for a bond indicates a larger bond dimension. (b) Magnitude of SCI
  coefficients $|c_i|$ (black dots) and truncation error $1-\sum_{j=0}^{i}|c_i|^2$ (red dots). (c) Bond dimension $D$ and von Neumann entropy $S_{\rvN}$ of MPS (topology A). (d) Sorted $D$ for three CTNS. (e) Sorted $S_{\rvN}$.}\label{fig:pn}
\end{figure*}

\begin{figure*}
  \begin{tabular}{cc}
  \multicolumn{2}{c}{\includegraphics[width=0.8\textwidth]{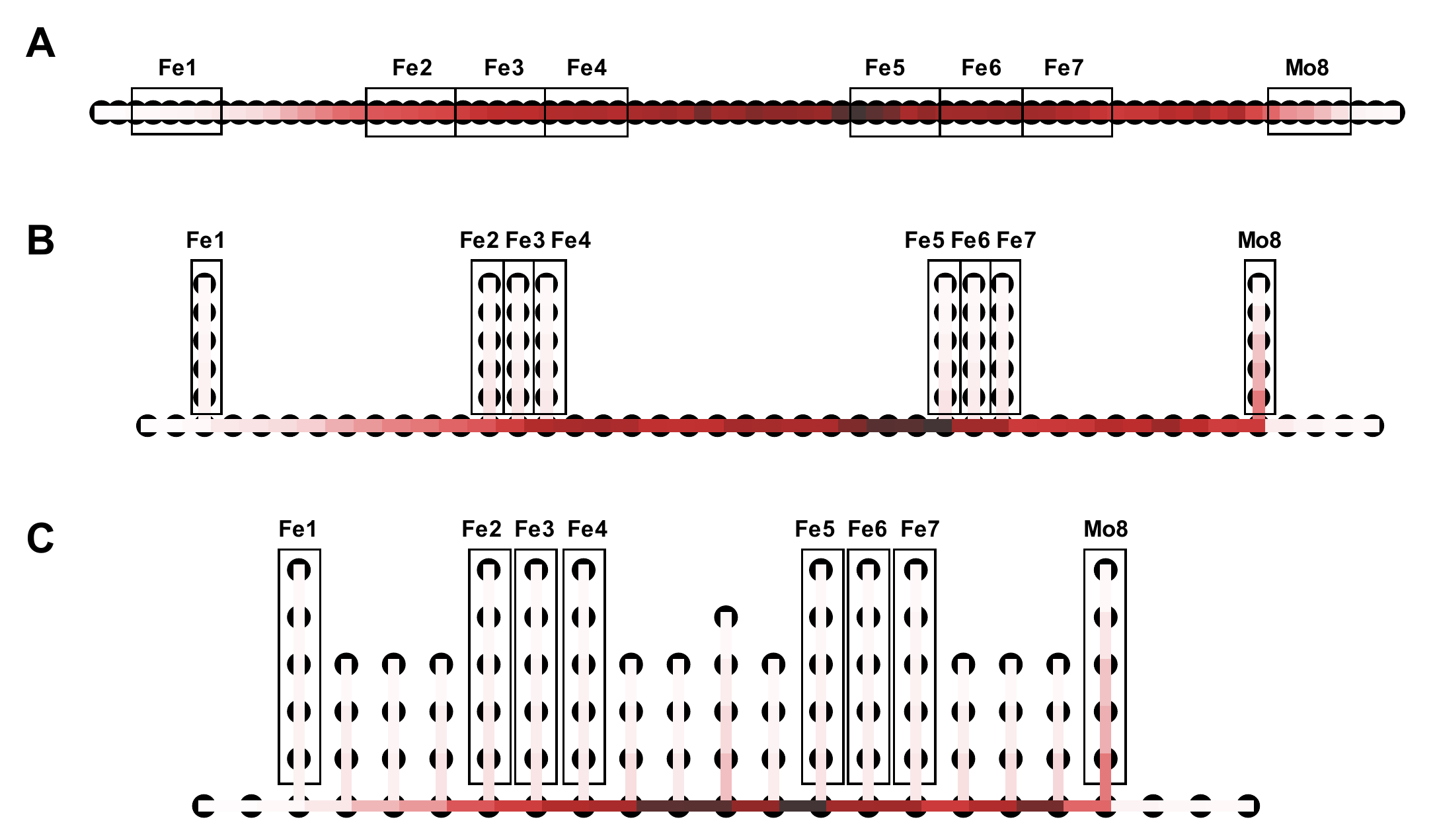}} \\
  \multicolumn{2}{c}{(a) three topologies of CTNS}\\
  \includegraphics[height=0.25\textheight]{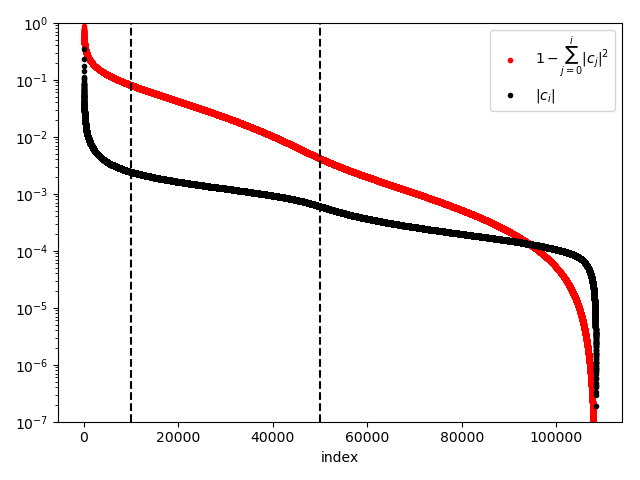} &
  \includegraphics[height=0.25\textheight]{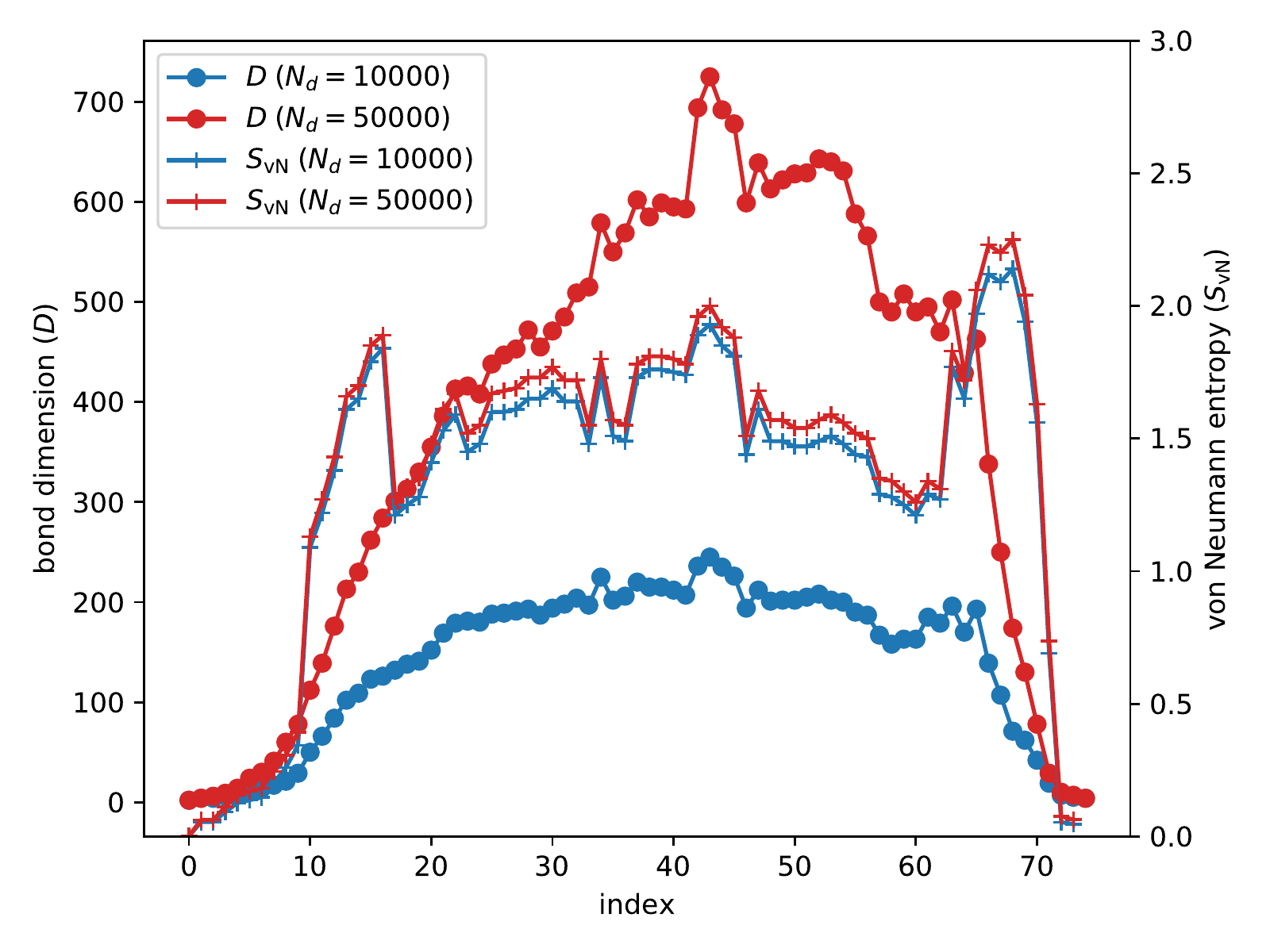} \\
  (b) magnitudes of SCI coefficients & (c) $D$ and $S_{\rvN}$ of MPS (topology A) \\
  \includegraphics[height=0.25\textheight]{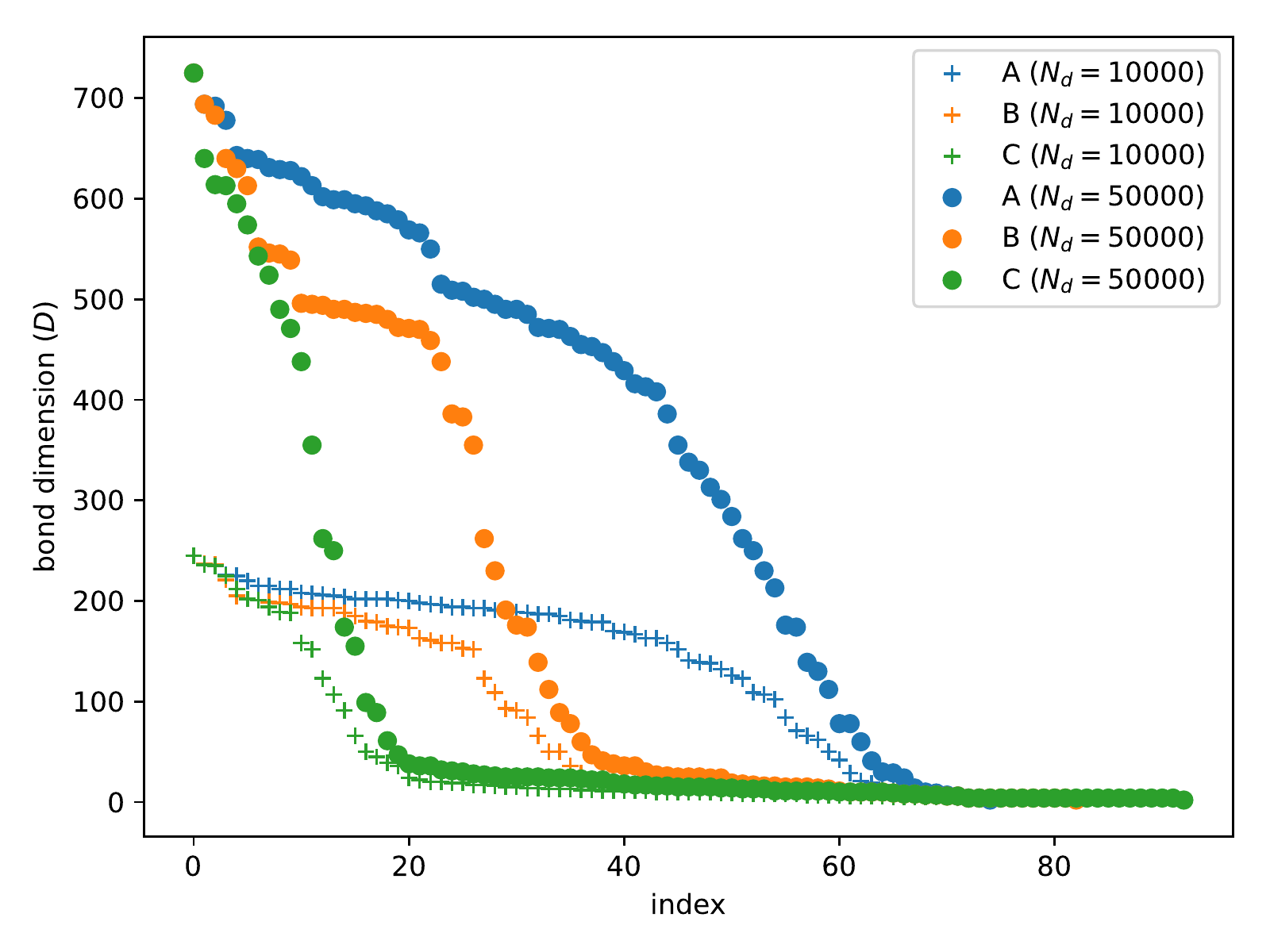} &
  \includegraphics[height=0.25\textheight]{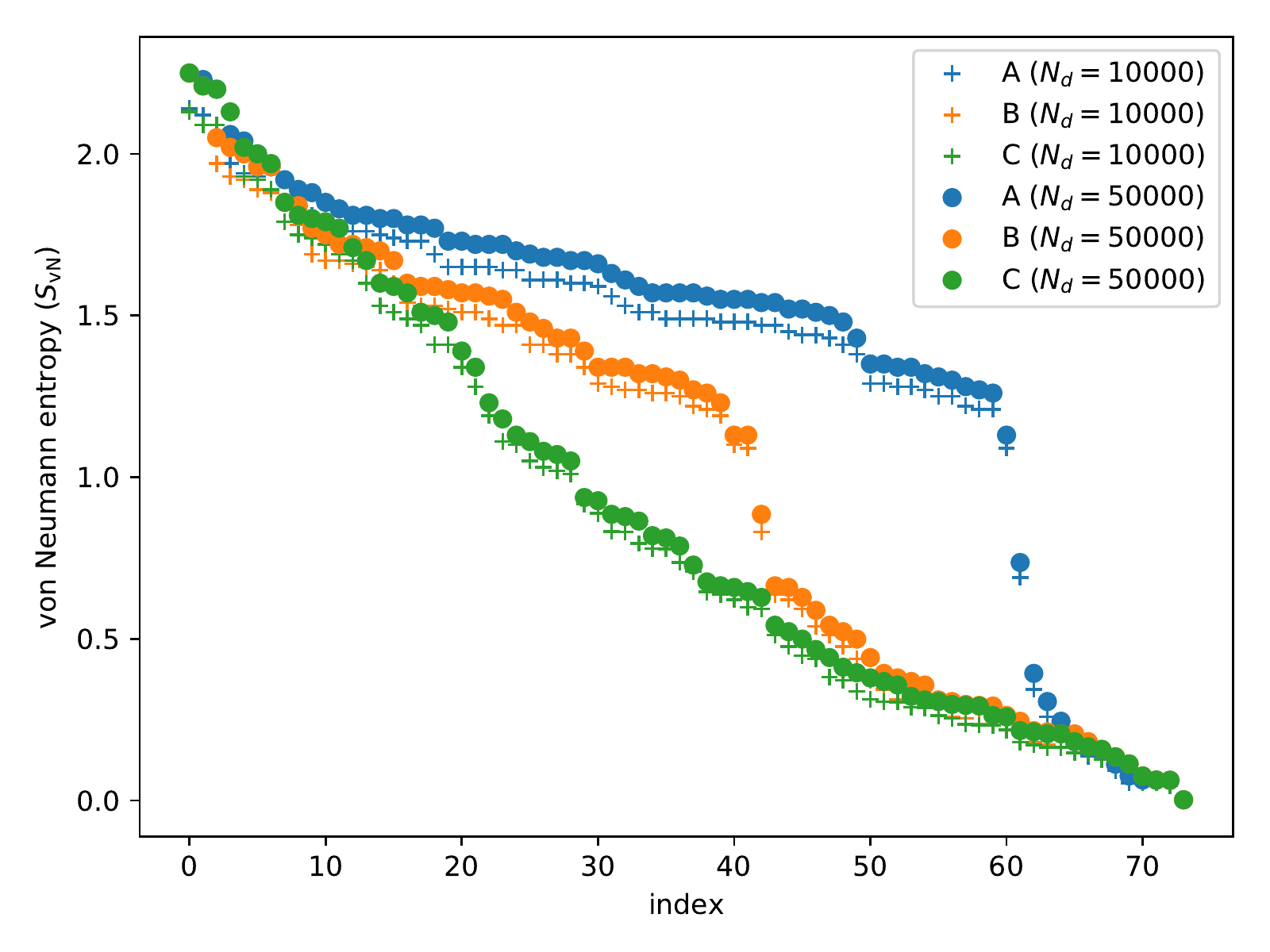} \\
  (d) sorted $D$ for three CTNS & (e) sorted $S_{\rvN}$
  \end{tabular}
  \caption{CTNS representations with different topologies of the SCI wavefunction ($M=3/2$) obtained with $\epsilon_1=10^{-3}$
  for the FeMo-cofactor in the active space CAS(113e,76o). The results obtained by retaining determinants with largest $N_d=10000$ and $N_d=50000$ magnitudes of SCI coefficients in the decomposition into CTNS representations are shown for comparison in (c,d,e).
  For other explanations, see Fig. \ref{fig:pn}.}\label{fig:femoco}
\end{figure*}

The computed results for the P-cluster and the FeMoco were summarized in Figs. \ref{fig:pn}
and \ref{fig:femoco}, respectively. Before discussing the expressibility of CTNS,
we should emphasize that
the quality of these SCI wavefunctions is
not expected to be high for such strongly correlated systems.
In fact, we found that the sampled determinants only correspond to a corner of the Hilbert space
around the initial broken-symmetry determinant. Even in this case,
as shown in Figs. \ref{fig:pn}(b)
and \ref{fig:femoco}(c), there are a large number of determinants
with magnitudes of coefficients around 10$^{-3}$, indicating
that using a truncated CI would not be sufficient for such systems.
Thus, we do not intend to draw conclusions about the nature of the true ground state based on these wavefunctions,
but just use them as representatives for multi-determinant wavefunctions to study the expressive power of CTNS
with different topologies.

For the P-cluster, as shown in Fig. \ref{fig:pn}(c), there is a valley along the MPS chain
for both bond dimensions $D$ and entanglement entropies $S_{\rvN}$, which suggests that
the left and right cubanes are entangled less strongly compared with the couplings
within each cubane. This is consistent with the geometry of the P-cluster
in its resting state, see Fig. \ref{fig:nitrogenase}(a), where
the two cubanes share a corner sulfide and are connected by two thiolate bridges.
In Figs. \ref{fig:pn}(d) and (e), the sorted $D$ and $S_{\rvN}$
are compared for the three different topologies. We can find that
both $D$ and $S_{\rvN}$ of topology C are grossly smaller than
those of topology A. A more detailed comparison can be seen
from Fig. \ref{fig:pn}(a), where the darkness of the color on each bond
represents the magnitude of the bond dimension. Clearly,
by putting the orbitals within an atom on the branches, as done
in topologies B and C, the number of strongly entangled sites on the backbone
is significantly reduced compared with that for the MPS chain.
However, it is important to realize that the entanglement entropy
at a given bond will be the same for different loop-free TNS,
if the corresponding bipartition of orbital space is the same.
Because in view of Eq. \eqref{eq:CIlr},
the singular values will be the same regardless
of the ordering of orbitals within each subspace.
This explains the observation of some coincidences of $D$ or $S_{\rvN}$
in Figs. \ref{fig:pn}(d) and (e).

For the FeMoco, we illustrated the decomposition for two truncated CI wavefunctions for comparison, obtained
by retaining determinants with largest $N_d=10000$ and $N_d=50000$ coefficients in magnitude,
respectively, from the computed SCI wavefunction. As shown in Fig. \ref{fig:femoco}(b),
the resulting CI wavefunctions correspond to roughly 90\% and 99\% fidelities, respectively.
We note that whereas a valley is observed for $D$ and $S_{\rvN}$ in the MPS chain for the P-cluster,
the same behavior is not observed in the corresponding Fig. \ref{fig:femoco}(c) for the FeMoco.
This suggests that the two cubanes in the FeMoco is more strongly entangled,
which seems to be reasonable considering the fact that
in the FeMoco the left cubane (Fe1, Fe2, Fe3, and Fe4) and
the right cubane (Fe5, Fe6, Fe7, and Mo8) are coupled in a face-to-face way
through the central carbon and three sulfide bridges (see Fig. \ref{fig:nitrogenase}(a)).
Future investigation of the elusive electronic structure of the FeMoco needs to be
carried out, once an efficient method has been developed.

As shown in Figs. \ref{fig:femoco}(c,d,e),
while the bond dimensions $D$ increase significantly for representing the CI wavefunction with
$N_d=50000$ faithfully, the entanglement entropies $S_{\rvN}$ do not increase too much
from $N_d=10000$ to $N_d=50000$. Similar to the P-cluster case, topologies B and C have grossly smaller $D$ and $S_{\rvN}$.
Thus, the CTNS with topology C may be a better variational ansatz for the FeMoco than
the simple MPS used previously\cite{li2019electronicFeMoco}.
However, whether the reduction of $D$ or $S_{\rvN}$ can turn
into a reduction of computational cost needs to be further investigated
in future. Because just consider solving a local CI problem
during the one-site sweep optimization, the computational cost
for the matrix-vector product in the Davidson diagonalization
scales as $O(K^2D^3)$ for MPS, and the total cost for $K$ sites
is $O(K^3D^3)$. In comparison, solving
a local CI problem for the internal sites
of CTNS scales as $O(K^2(D_1^3D_2+D_1^2D_2^2))$,
where $D_1$ ($D_2$) represents the bond dimension
on the backbone (branches), while that for the other
sites of CTNS scales as $O(K^2D_1^3)$ or $O(K^2D_2^3)$
depending on whether the site is on the backbone or branches.
Therefore, if $D_1=D$ then the local problem
for optimizing the internal sites of CTNS will be
more expensive than that for DMRG, formally by a factor of $D_2$.
However, we hope that by making the backbone shorter via
introducing short branches, $D_2$ can be made
small and the total computational cost can be reduced
if the length of the backbone is much smaller than $K$.
In practice, topologies similar to topology B with a small number of internal sites
may be a good candidate for efficient ansatz.
This will be the subject of our future research.

\section{Conclusion and outlook}\label{sec:conclusion}
In this work, we proposed the use of CTNS for tackling
strongly correlated polynuclear transition metal compounds,
which can be viewed as an effective coarse-graining
approach to compactly describe both the intra-atomic
and interatomic electron correlations.
As the first step, the expressibility of CTNS was investigated for the
P-cluster and the FeMoco of nitrogenase using approximate CI wavefunctions
generated from SCI calculations as representatives.
It is shown that compared with MPS, the bond dimensions necessary
to represent the same SCI wavefunction are significantly reduced
in CTNS with a chemically more meaningful topology for these challenging
clusters. However, whether this reduction can transform into
computational advantages is an intriguing open question,
considering the fact that the generic CTNS are more complex than MPS.
Work in this direction is currently being carried out.
A pilot implementation of the DMRG-like sweep algorithm using
the complementary operator approach has been made for variationally optimizing CTNS.
A more efficient implementation with parallelization needs to be developed in order to
make a fair comparison with the state-of-the-art implementation
of DMRG. Overall, we suggest CTNS as a promising class of TNS
for studying electronic structures of polynuclear transition metal compounds.

\section*{Appendix: Details of the three topologies of CTNS}
We documented the details of the three topologies of CTNS used in this work
with the previously reported active space models\cite{li2019electronicPcluster,li2019electronicFeMoco} of the P-cluster and the FeMoco.
The active orbitals were ordered by the genetic
ordering method\cite{olivares-amaya_ab-initio_2015} for MPS.
Numbers in each parenthesis represent the indices of molecular orbitals within a branch of CTNS.

P-cluster:
\begin{enumerate}
\item topology A:
(0), (1), (2), (3), (4), (5), (6), (7), (8), (9), (10), (11), (12), \
(13), (14), (15), (16), (17), (18), (19), (20), (21), (22), (23), \
(24), (25), (26), (27), (28), (29), (30), (31), (32), (33), (34), \
(35), (36), (37), (38), (39), (40), (41), (42), (43), (44), (45), \
(46), (47), (48), (49), (50), (51), (52), (53), (54), (55), (56), \
(57), (58), (59), (60), (61), (62), (63), (64), (65), (66), (67), \
(68), (69), (70), (71), (72)

\item topology B:
(2), (0), (1), (8), (3, 4, 5, 6, 7), (9), (10), (11), (12),
 (20), (13), (14), (22), (21),
 (15, 16, 17, 18, 19),
 (23, 24, 25, 26, 27),
 (28, 29, 30, 31, 32),
 (34), (33), (35), (36), (37), (39), (38),
 (40, 41, 42, 43, 44),
 (45, 46, 47, 48, 49),
 (50), (51), (52), (54),
 (59, 60, 61, 62, 63), (53), (56), (55), (58), (57), (64, 65, 66, 67,
  68),
 (69), (70), (71), (72)

\item topology C:
(2),
 (8),
 (0, 1),
 (3, 4, 5, 6, 7),
 (9, 12, 14),
 (10, 11, 13),
 (15, 16, 17, 18, 19),
 (20, 21, 22),
 (23, 24, 25, 26, 27),
 (28, 29, 30, 31, 32),
 (33, 34, 37),
 (35, 39),
 (36, 38),
 (40, 41, 42, 43, 44),
 (45, 46, 47, 48, 49),
 (50, 51, 52),
 (54, 56, 58),
 (53, 55, 57),
 (59, 60, 61, 62, 63),
 (64, 65, 66, 67, 68),
 (69, 70),
 (71),
 (72)
\end{enumerate}

FeMoco:
\begin{enumerate}
\item topology A:
(0), (1), (2), (3), (4), (5), (6), (7), (8), (9), (10), (11), (12), \
(13), (14), (15), (16), (17), (18), (19), (20), (21), (22), (23), \
(24), (25), (26), (27), (28), (29), (30), (31), (32), (33), (34), \
(35), (36), (37), (38), (39), (40), (41), (42), (43), (44), (45), \
(46), (47), (48), (49), (50), (51), (52), (53), (54), (55), (56), \
(57), (58), (59), (60), (61), (62), (63), (64), (66), (67), (68), \
(65), (69), (70), (71), (72), (73), (74), (75)

\item topology B:
(0), (1), (2, 3, 4, 5, 6),
 (7), (8), (9), (10), (11), (12), (14), (13), (15),
 (16, 17, 18, 19, 20),
 (21, 22, 23, 24, 25),
 (26, 27, 28, 29, 30),
 (31), (33), (35), (41), (32), (37), (36), (38), (40), (42), (34), \
(39), (43),
 (44, 45, 46, 47, 48),
 (49, 50, 51, 52, 53),
 (54, 55, 56, 57, 58),
 (60), (59), (61), (62), (68), (63), (64), (66),
 (65, 69, 70, 71, 72),
 (67),
 (73),
 (74),
 (75)

\item topology C:
(0), (1),
 (2, 3, 4, 5, 6),
 (8, 13, 15),
 (7, 12, 14),
 (9, 10, 11),
 (16, 17, 18, 19, 20),
 (21, 22, 23, 24, 25),
 (26, 27, 28, 29, 30),
 (31, 32, 34),
 (33, 39, 40),
 (35, 36, 37, 38),
 (41, 42, 43),
 (44, 45, 46, 47, 48),
 (49, 50, 51, 52, 53),
 (54, 55, 56, 57, 58),
 (59, 60, 63),
 (61, 62, 67),
 (64, 66, 68),
 (65, 69, 70, 71, 72),
 (73),
 (74),
 (75)
\end{enumerate}

\section*{Acknowledgements}
This work was supported by the National Natural Science Foundation of China (Grants
No. 21973003) and the Beijing Normal University Startup Package.

\bibliographystyle{apsrev4-1}
\bibliography{references}

\end{document}